\RequirePackage[2020-02-02]{latexrelease}
\documentclass[aps,preprint,11pt]{revtex4}
\usepackage{amsmath}
\usepackage{graphicx}
\usepackage{amsfonts}
\usepackage{amssymb}
\usepackage{blindtext}
\usepackage{subcaption}
\usepackage[T1]{fontenc}
\usepackage[utf8]{inputenc}
\usepackage{float}
\usepackage{xcolor}
\DeclareMathOperator{\sech}{sech}

\RequirePackage[2020-02-02]{latexrelease}
\RequirePackage[2020-02-02]{latexrelease}
\providecommand{\U}[1]{\protect\rule{.1in}{.1in}}
\providecommand{\U}[1]{\protect\rule{.1in}{.1in}}
\providecommand{\U}[1]{\protect\rule{.1in}{.1in}}
\providecommand{\U}[1]{\protect\rule{.1in}{.1in}}
\providecommand{\U}[1]{\protect\rule{.1in}{.1in}}

\begin{document}
	
	\title{One dimensional Bose-Einstein condensate under the effect of the extended uncertainty principle}
	\author{Abderrahmane Benhadjira$^1$}
	\email{benhadjira.abdelrahman@univ-ouargla.dz}
 \email{abenhadjira97@gmail.com}
 
	\author{Abdelhakim Benkrane$^2$}
	\email{abdelhakim.benkrane@univ-ouargla.dz}
 \email{hakim9502.benkrane@gmail.com}
	\author{Omar Bentouila$^1$}
	\author{Hadjira Benzair$^2$}
	\author{Kamal Eddine Aiadi$^1$}
	
	\affiliation{$^1$Équipe Optoélectronique, LENREZA Laboratory, University of Kasdi Merbah, Ouargla, 30000, Algeria}
	\affiliation{$^2$Université Kasdi Merbah Ouargla, Laboratoire LRPPS, Ouargla 30000, Algeria}
	
	\begin{abstract}
 		In this study, an analytical investigation was conducted to assess the effects of the extended uncertainty principle (EUP) on a Bose-Einstein condensate (BEC) described by the deformed one-dimensional Gross-Pitaevskii equation (GPE). Analytical solutions were derived for null potential while we used variational and numerical methods for a harmonic oscillator potential. The effects of EUP on stability, probability density, position, and momentum uncertainties of BEC are analyzed. The EUP is found to be applicable for the free dark soliton solution and in the presence of a harmonic potential within specific ranges of the deformation parameter $\alpha$, while it is not valid for the free bright soliton solution.
	\end{abstract}
	
	\maketitle
	
	\section{Introduction}
The Gross-Pitaevskii equation (GPE) is fundamentally used to describe the behavior of a Bose-Einstein condensate (BEC) at very low temperatures ($T \rightarrow 0, \mathrm{K}$) \cite{31,32}. Originally, it appeared in the beginning as a model to study superfluids and the vortex lines modeled by a non-perfect Bose-Einstein gas \cite{GPE, GPE1, GPE2}. The GPE is a nonlinear Schrödinger equation with cubic nonlinearity. It has been proven that the 1D cubic nonlinear Schrödinger equation satisfies a solitonic (shape-invariant) solution \cite{GPE4}, specifically, bright and dark soliton solutions for self-focusing and self-defocusing respectively.
Several works have been done in the context of addressing the GPE experimentally, analytically, and numerically under various types of nonlinearity and potentials. To name a few, A. Weller et al. \cite{GPE5} experimentally investigated the oscillating and interacting matter wave dark solitons by releasing a BEC from a double well potential into a harmonic trap in the crossover regime between one dimension and three dimensions. Similarly, A.D Carli et al. \cite{GPE6} experimentally studied the excitation modes of bright matter-wave solitons created by quenching the interactions of a BEC of cesium atoms in a quasi-one-dimensional geometry. Furthermore, B. Gertjerenken et al. \cite{GPE7} studied analytically and numerically the bright solitons scattering of potential in a one-dimensional geometry via the GPE. L. Salasnich et al. \cite{GPE8} numerically investigated the formation of bright solitons in a BEC using the GPE with a dissipative three-body term.

In recent years, the extended uncertainty principle (EUP) has attracted great interest due to its ability to replicate the effects of space-time curvature \cite{EPJ2020}. Furthermore, it is important to study quantum effects over large distances \cite{Physlett}. In this context, numerous applications have emerged, including the one-dimensional Klein-Gordan and Dirac oscillators \cite{Merad2019}, two-dimensional Dirac oscillator \cite{Benkrane}, three-dimensional relativistic Coulomb potential \cite{Hamil2021}, the two-dimensional Dirac equation with Aharonov–Bohm-Coulomb interaction \cite{Hamilscripta}, and scalar and vector potentials for the Klein-Gordon equation in one dimension \cite{A.Merad2019}.
	
	Our goal is to study the one-dimensional  GPE for the bright and dark solitons in the presence of EUP. For this purpose, we generalize the momentum operator as follows
	\begin{equation}
		\hat{P}=-i\hbar f_{\alpha}(x) \frac{d}{dx}. \label{1}
	\end{equation}
The EUP is encoded in the positive function $f_{\alpha}(x)$  with $\alpha$ a small and positive parameter, where $f_{\alpha=0}(x)=1$.  
 	
	 Eq. (\ref{1}) leads to the following modified commutation relation and uncertainty principle
	\begin{gather}
		[\hat{x},\hat{P}]=i\hbar f_{\alpha}(x)\\
  \Delta X \Delta P\geq \dfrac{\hbar}{2} \big< f_{\alpha}(x)\big>,
 \end{gather}
$\big< f_{\alpha}(x)\big>$ being the mean value of $f_{\alpha}(x)$. The momentum operator in Eq. (\ref{1}) is not Hermitian anymore. Therefore the Hamiltonian of the system is not Hermitian as well, so the operations, such as
	the scalar product between two functions and  the closure and projection relations must be modified respectively as follows:
	\begin{equation}
		\left\langle \Psi\mid \Phi\right\rangle _{\alpha}=\int\frac{dx}{f_{\alpha}(x)}\Psi^{\star}(x)\Phi(x),\label{5}%
	\end{equation}
	\begin{equation}
		\int\frac{dx}{f_{\alpha}(x)}\lvert x \rangle\langle x \lvert
		=\mathbb{I},\label{6}%
	\end{equation}
	\begin{equation}
		\langle y\lvert z\rangle_{\alpha}=f_{\alpha}(x) \delta(y-z).\label{7}%
	\end{equation}
In the context of general relativity, $1/(f_{\alpha}(x))^2$ can represent the spatial  component of  the metric tensor for $(1+1)$-dimensional curved space-time; $(f_{\alpha}(x))^{-2}=g_{xx}$ with determinant $1/f_{\alpha}(x)$ \cite{costafilho, gine}. Therefore, as we mentioned earlier,  the impact of the EUP is analogous to the curvature of space-time and, consequently, has a strong connection to the gravitational fields. The investigation of the influence of the EUP on the BEC has been addressed in Ref. \cite{velocity}, where, the authors consider the effects originating from the Planck scale regime during the free expansion of a condensate. They naturally introduce a modified uncertainty principle, which is associated with the quantum structure of space-time. Additionally, in Ref \cite{Qdeformed1}, the authors examine the GPE within the framework of $q$-deformed or nonextensive statistics which is mathematically equivalent to Eq. (\ref{1}). This approach aligns with experimental observations of anyons \cite{science1, science2, science3}.
	To our best knowledge, the topic of the GPE  has been addressed in terms of a modified momentum operator only once  \cite{Qdeformed1}.
 
 In the current study, we aim to investigate the influence the impact of the EUP by using Eq. (\ref{1}). Our focus will be on analyzing the probability densities, position, and momentum uncertainties of a BEC in both null and harmonic trap potentials.

		\section{Mathematical analysis}
	The simplest form  of the  ordinary one-dimensional GPE ($\hbar=m=1$ )\cite{2013} is given by
	\begin{gather}
		i\frac{\partial \Psi(x,t)}{\partial t}=-\frac{1}{2}\frac{\partial^{2} \Psi(x,t)}{\partial x^{2}}+g\left\arrowvert  \Psi(x,t) \right\arrowvert^{2} \Psi(x,t)+V(x)  \Psi(x,t), \label{vbb}
	\end{gather}	
	Where, $x$ represents the position and $-\infty < x < +\infty$,  $V(x)$ is the scalar potential of the system, $ \Psi(x,t)$ a complex wave function where $\left\arrowvert  \Psi(x,t) \right\arrowvert^{2}$ is interpreted as the atomic density \cite{2013,salazar}, and $g$ a parameter that determines the type of interaction between atoms \cite{salazar}. $g>0$ and $g<0$  are for repulsive and attractive interactions respectively \cite{2013}.
	
In the context of the EUP, Eq. (\ref{vbb}) becomes: 
	\begin{gather}
		i\frac{\partial \Psi(x,t)}{\partial t}=-\frac{1}{2}f_{\alpha}(x)\left(\frac{\partial }{\partial x}f_{\alpha}(x)\frac{\partial}{\partial x}\right) \Psi(x,t)+g\left\arrowvert  \Psi(x,t) \right\arrowvert^{2}\Psi(x,t) +V(x)  \Psi(x,t). \label{vbbb}
	\end{gather}
The existence of a generalized momentum operator leads to an unusual form of the kinetic energy operator, and it may create an effective potential different from the original potential.
 By using the transformation $\displaystyle d\eta=\frac{dx}{f(x)}$, and  putting 
						$\Psi(x,t)=\tilde{\Psi}(\eta,t)$, Eq. (\ref{vbbb}) transforms to
						\begin{gather}
							i\frac{\partial \tilde{\Psi}(\eta,t)}{\partial t}=-\frac{1}{2}\frac{\partial^{2} \tilde{\Psi}(\eta,t)}{\partial \eta^{2}}+g\left|  \tilde{\Psi}(\eta,t) \right| ^{2}\tilde{\Psi}(\eta,t)   +V(x(\eta))\tilde{\Psi}(\eta,t). \label{vbbb2}
						\end{gather}
						Since we are going to focus on stationary systems, the general solution can be written as
						\begin{equation}
							\tilde{\Psi}(\eta,t)=\tilde{\Phi}(\eta)\exp(-i\varepsilon t),
						\end{equation}
						where $\varepsilon$ can be interpreted as a possible energy level or as the chemical potential of the system. 
	
	Throughout this study, we consider the deformation function $f_{\alpha}(x)=1+\alpha x^2$,  which leads to the following EUP
 \begin{gather}
     \Delta X \Delta P\geq \dfrac{1}{2}\left(1+\alpha (\Delta X)^{2}\right).\label{euppp}
 \end{gather}
This implies a maximal length $(\Delta X)_{\text{max}}=1/\sqrt{\alpha}$ and a minimal momentum $(\Delta P)_{\text{min}}=\sqrt{\alpha}$. The relation (\ref{euppp})  simulates an anti-de Sitter space \cite{Migenmi,Bolen}. Under these considerations, the new variable $\eta$ is expressed as $\eta=\frac{\arctan(x\sqrt{\alpha})}{\sqrt{\alpha}}$. This transformation allows us to obtain the familiar expressions for the normalization condition and the kinetic energy operator.
	\subsection{Null potential [V(x)=0]}
	For $V(x)=0$ and $g=-1$, Eq. (\ref{vbb}) exhibits a bright soliton which can be written as \cite{2013}
	\begin{gather}
		\Psi(x,t)=\sqrt{2\mu}\sech(\sqrt{2\mu} x)e^{i\mu t},\label{g}
	\end{gather}	
	with $\mu$ the chemical potential. On the other hand for $g=+1$, it yields a dark soliton solution given by \cite{2016}
	\begin{gather}
		\Psi(x,t)=\sqrt{\mu}\tanh(\sqrt{\mu}x)e^{-i\mu t}.\label{gh}
	\end{gather}
	For simplicity, we have assumed the following:\\ 
 (i) The bright soliton is at rest,  (ii)   the initial location of the dark soliton is $x_0=0$, (iii) the background of the dark soliton is stationary,  (vi) the relative velocity between the dark soliton and the background is null. 
 Under the effect of EUP, the solutions (\ref{g}) and (\ref{gh}) become
	\begin{gather}
		\tilde{\Psi}_{g=-1}(\eta,t)=\sqrt{2\mu}\sech(\sqrt{2\mu}\eta)e^{i\mu t}, \label{l}
	\end{gather}
	\begin{gather}
		\tilde{\Psi}_{g=+1}(\eta,t)=\sqrt{\mu}\tanh(\sqrt{\mu}\eta)e^{-i\mu t}.  \label{ll}
	\end{gather}
 For the chosen EUP (\ref{euppp}), the bright and dark soliton solutions in terms of the original variable $x$ are expressed as
	\begin{gather}
		\Psi_{g=-1}(x,t)=\sqrt{2\mu}\sech\left(\sqrt{2\mu} \frac{\arctan(\sqrt{\alpha} x)}{\sqrt{\alpha}}\right)e^{-i\mu t},\label{g1}
	\end{gather}
	and 
	\begin{gather}
		\Psi_{g=1}(x,t)=\sqrt{\mu}\tanh\left(\sqrt{\mu}\frac{\arctan(\sqrt{\alpha} x)}{\sqrt{\alpha}}\right)e^{-i\mu t},
        \label{gh1}
	\end{gather}
	respectively.  For the bright soliton case, the expectation  value of $\hat{X^{2}}$ diverges
 \begin{gather}
      \big <\hat{X}^{2}\big>=\int_{-\infty}^{+\infty}\dfrac{2\mu\sech^{2}\left(\sqrt{2\mu} \frac{\arctan(\sqrt{\alpha} x)}{\sqrt{\alpha}}\right) x^{2}}{1+\alpha x^{2}}dx 	\rightarrow +\infty.
 \end{gather}
However, the EUP leads to a bounded dense domain of the infinite-dimensional Hilbert space $\mathcal{H}$ due to the maximum length constraint \cite{Lawson}. Consequently, instead of studying the entire Hilbert space $\mathcal{H}=\mathcal{L}^{2}(\mathbb{R})$, we investigate a limited subspace  $\mathcal{D}_{\alpha}=\mathcal{L}^{2}(a,b)$ where $ a,b \in \mathbb{R}$ and take the values $a=-\ell_{\text{max}}=-1/\sqrt{\alpha}, b=\ell_{\text{max}}=1/\sqrt{\alpha}$. In the case of the usual quantum mechanics where $\alpha=0$, we recover the total Hilbert space $ \mathcal{D}_{\alpha=0}=\mathcal{L}^{2}(\mathbb{R})$. It is worth mentioning that the deformation in momentum space, also known as the generalized uncertainty principle (GUP), leads to minimal length constraint. Additional details on this aspect can be found in \cite{ReviewD}.

The normalization of the solutions (\ref{l}) and (\ref{g1}) leads to the following transcendental relation between $\alpha$ and $\mu$
	\begin{gather}
		2\sqrt{2\mu}\tanh\left(\pi\sqrt{\frac{\mu}{8\alpha}}\right)=1. \label{norm}
	\end{gather}
In the conventional case where $\alpha=0$, the chemical potential takes the value $\displaystyle\mu= 1/8$. Additionally, it is crucial to highlight that as a result of Eq.(\ref{5}), the probability density is expressed as $\left|\Psi(x,t) \right|^{2}/(1+\alpha x^{2})$ instead of just $\left|\Psi(x,t) \right|^{2}$.

Further, the uncertainty of a quantity $\hat{Q}$ is given by
 \begin{gather}
     \Delta Q=\sqrt{\big<\hat{Q}^{2}\big>-\big<Q\big>^{2}},
 \end{gather}
where $\big<\hat{Q}\big>$ is the mean value of the operator $\hat{Q}$.  One can easily see that  the bright soliton solution is an even function; therefore, the mean value of the position is $\langle\hat{X}\rangle=0$. On the other hand, the mean value of $\hat{X}^{2}$ is calculated as
 \begin{equation}
    \big <\hat{X}^{2}\big>=\int_{-\ell_{max}}^{+\ell_{max}}\dfrac{2\mu\sech^{2}\left(\sqrt{2\mu} \frac{\arctan(\sqrt{\alpha} x)}{\sqrt{\alpha}}\right) x^{2}}{1+\alpha x^{2}}dx.
 \end{equation}
  Where we used the modified scalar product expressed in Eq. (\ref{5}). Hence, the position  uncertainty $\Delta X$ takes the form
	\begin{gather}
		\Delta X=\sqrt{\int_{-\ell_{max}}^{+\ell_{max}}\dfrac{2\mu\sech^{2}\left(\sqrt{2\mu} \frac{\arctan(\sqrt{\alpha} x)}{\sqrt{\alpha}}\right) x^{2}}{1+\alpha x^{2}}dx}. \label{unced}
	\end{gather}
$\Delta X$  (\ref{unced}) can be expressed in a simpler form in terms of the new variable $\eta$ as 
 \begin{gather}
     \Delta X=\sqrt{\int_{-\pi/(4\sqrt{\alpha})}^{\pi/(4\sqrt{\alpha})} \frac{2\mu\tan^{2}(\sqrt{\alpha} \eta)\sech^{2}(\sqrt{2\mu}\eta)}{\alpha} d\eta}.\label{cc}
 \end{gather}

Similarly for the momentum uncertainty $\Delta P$, we employ $\hat{P}=-i (1+\alpha x^{2})\partial/\partial x=-i\partial/\partial \eta$ and $\Delta P$ becomes 

\begin{gather}
    \Delta P=\sqrt{-\int_{-\pi/(4\sqrt{\alpha})}^{\pi/(4\sqrt{\alpha})} 2\mu\sech(\sqrt{2\mu}\eta)\dfrac{\partial^{2}}{\partial \eta^{2}}\sech(\sqrt{2\mu}\eta) d\eta}. \label{momen}
\end{gather}

\begin{figure}
    \centering

    \begin{subfigure}{0.49\textwidth}
        \centering
        \includegraphics[width=\textwidth]{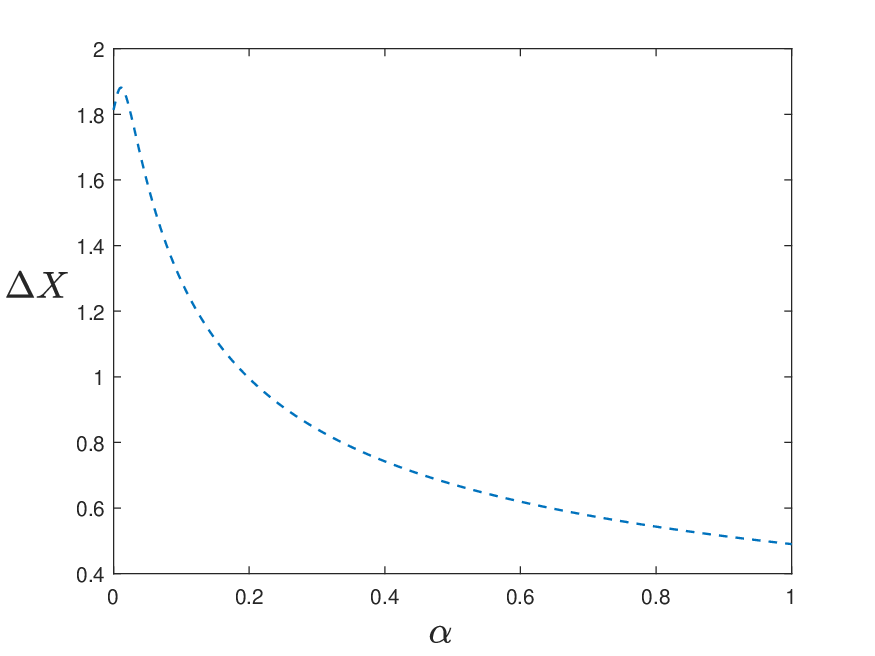}
        \caption{}
    \end{subfigure}
    \hfill
    \begin{subfigure}{0.49\textwidth}
        \centering
        \includegraphics[width=\textwidth]{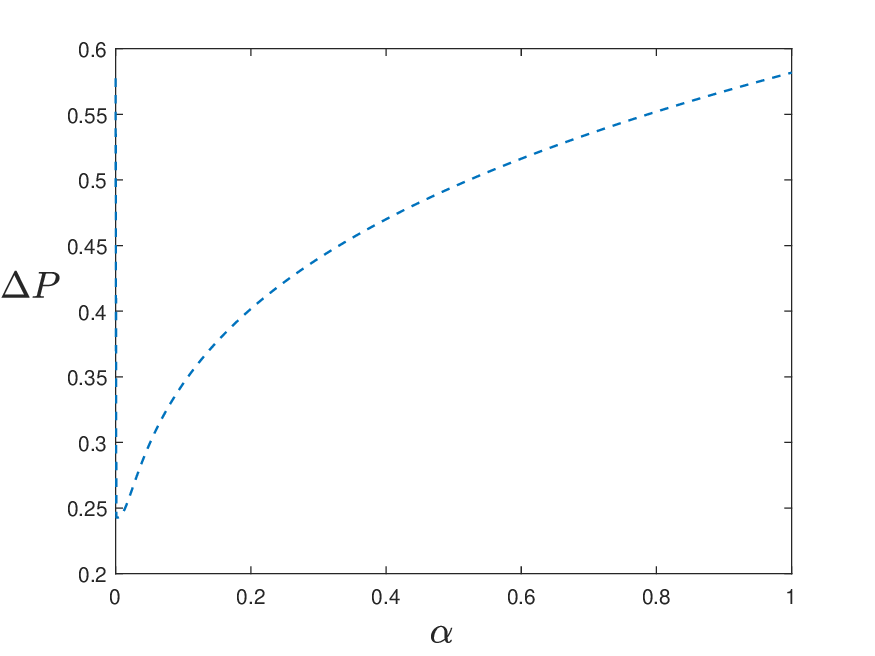}
        \caption{}
    \end{subfigure}

    \begin{subfigure}{0.5\textwidth}
        \centering
        \includegraphics[width=\textwidth]{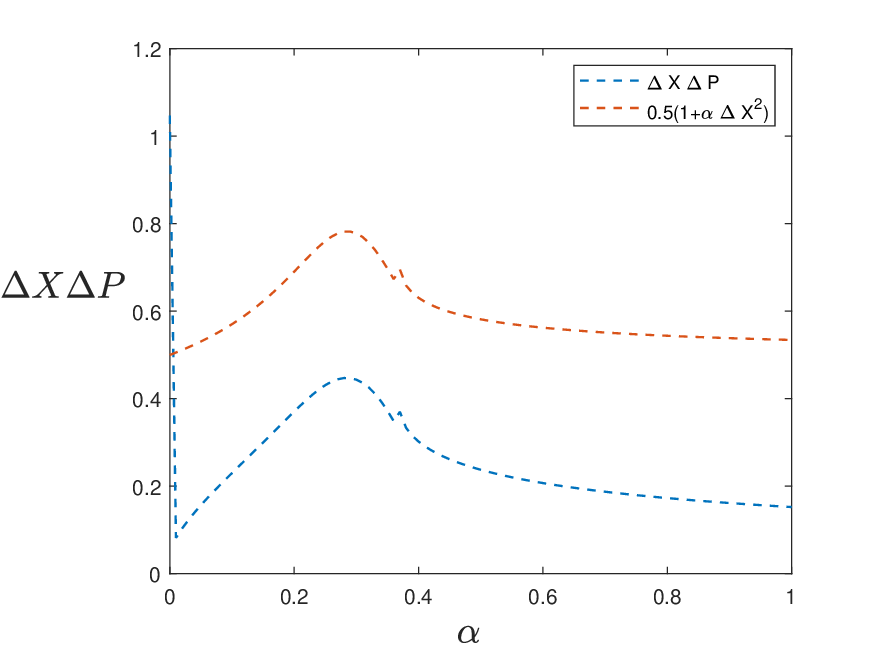}
        \caption{}
    \end{subfigure}

    \caption{Bright soliton solution: (a) $\Delta X$ vs $\alpha$ [Eq. (\ref{cc})]; (b) $\Delta P$ vs $\alpha$ [Eq. (\ref{momen})] \ ; (c) EUP validity: $\Delta X \Delta P$ against $0.5(1+\alpha \Delta X^2)$ [Inequality.(\ref{euppp})]}
    \label{freebright}
\end{figure}

As we can see from the Eqs. (\ref{cc}) and (\ref{momen}), $\Delta X $ and $\Delta P$ are influenced by $\alpha$. To visualize this impact, we depict in terms of $\alpha$ the variation of $\Delta X$, $\Delta P$, and the validation of the EUP in Fig. \ref{freebright}.

\begin{figure} [H]
    \centering

    \begin{subfigure}{0.49\textwidth}
        \centering
        \includegraphics[width=\linewidth]{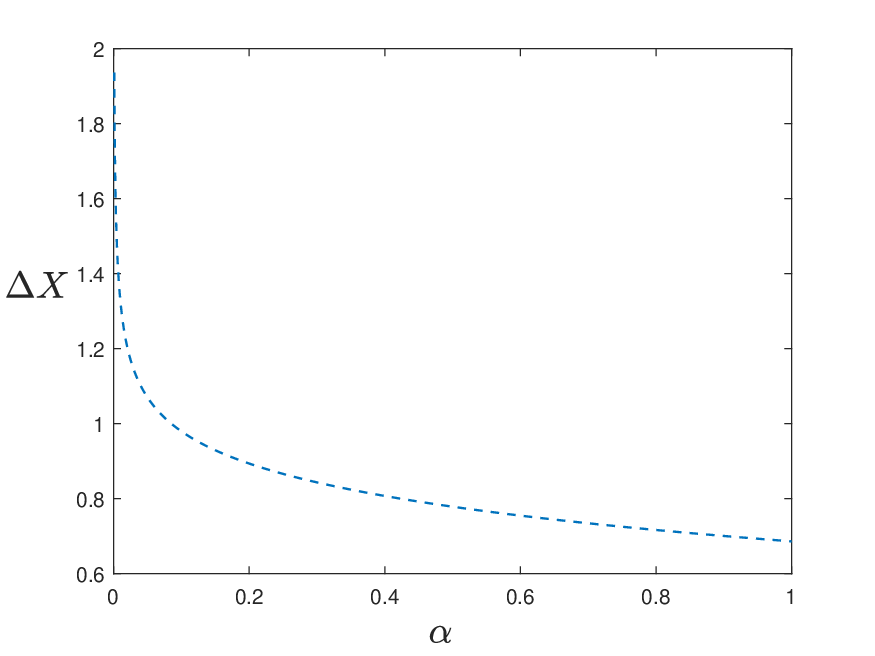}
        \caption{}
        \label{darkxun}
    \end{subfigure}
    \hfill
    \begin{subfigure}{0.49\textwidth}
        \centering
        \includegraphics[width=\linewidth]{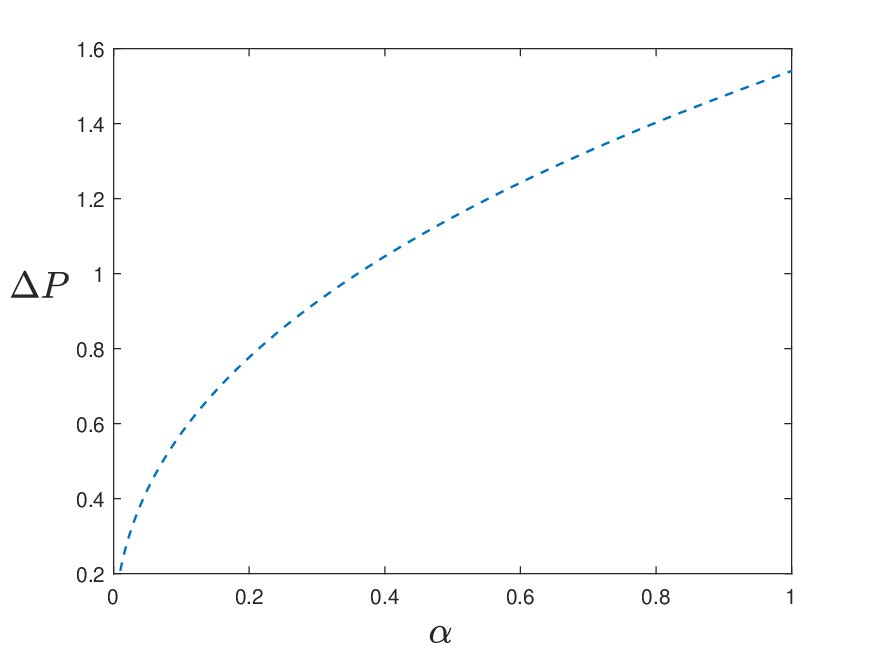}
        \caption{}
    \end{subfigure}

    \begin{subfigure}{0.49\textwidth}
        \centering
        \includegraphics[width=\linewidth]{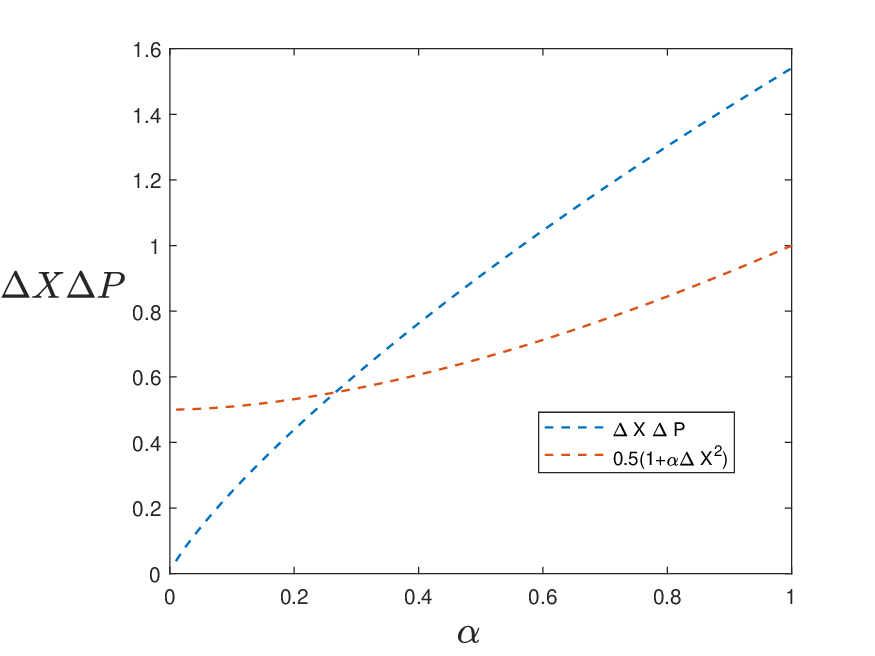}
        \caption{}
        \label{eupdark}
    \end{subfigure}
    \caption{ Dark soliton solution: (a) $\Delta X$ vs $\alpha$ [Eq. (\ref{cc})]; (b) $\Delta P$ vs $\alpha$ [Eq. (\ref{momen})] \ ; (c) EUP validity: $\Delta X \Delta P$ against $0.5(1+\alpha \Delta X^2)$ [Inequality. (\ref{euppp})]}
    \label{freedark}
\end{figure}

\begin{figure}[H]
	\centering	
	\includegraphics[width=0.7\textwidth]{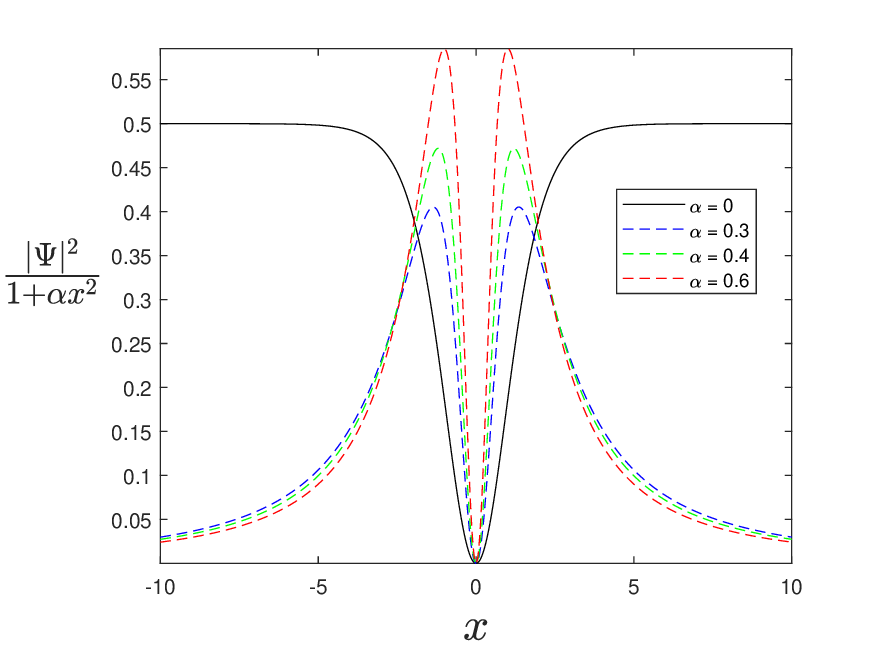} \caption{Squared norm of dark soliton solution and probability density distributions corresponding to the wavefunction described by Eq. (\ref{gh1}) for different values of $\alpha$ } 
		\label{darksoliton}%
	\end{figure}

For the dark soliton, unlike the ordinary case, the wave function can be normalized under the effect of the EUP. Consequently, we can establish a transcendental relation between the  $\alpha$ and the chemical potential $\mu$ which reads
	\begin{equation}
		\frac{\mu\pi}{2\sqrt{\alpha}}-2\sqrt{\mu}\tanh\left(\frac{\sqrt{\mu}\pi}{4\sqrt{\alpha}}\right)=1.
	\end{equation}
$\Delta X $ and $\Delta P$ of the dark soliton solution can be calculated respectively as follows:
\begin{gather}
     \Delta X=\sqrt{\int_{-\pi/(4\sqrt{\alpha})}^{\pi/(4\sqrt{\alpha})} \frac{\mu\tan^{2}(\sqrt{\alpha} \eta)\tanh^{2}(\sqrt{\mu}\eta)}{\alpha} d\eta} \label{xdark}
 \end{gather}

\begin{gather}
    \Delta P=\sqrt{-\int_{-\pi/(4\sqrt{\alpha})}^{\pi/(4\sqrt{\alpha})} \mu\tanh(\sqrt{\mu}\eta)\dfrac{\partial^{2}}{\partial \eta^{2}}\tanh(\sqrt{\mu}\eta) d\eta}. \label{pdark}
\end{gather}
$\Delta X$ and $\Delta P$ along with the validation of EUP for the dark soliton solution are depicted in Fig. \ref{freedark}. Additionally in Fig. \ref{darksoliton} we have plotted the probability density.

\subsection{Harmonic trap} 
In the presence of an external potential, obtaining an exact solution to the GPE becomes intractable and challenging due to its nonlinear nature. Thus, the focus of this section is to derive a qualitative solution for Eq. (\ref{vbbb}) subjected to a harmonic oscillator (HO) potential using the variational method and validate this latter through a numerical technique namely the split-step Fourier method (SSFM).
\subsubsection{Variational analysis}
\begin{figure}[H]
    \centering

    \begin{subfigure}{0.49\textwidth}
        \centering
        \includegraphics[width=\linewidth]{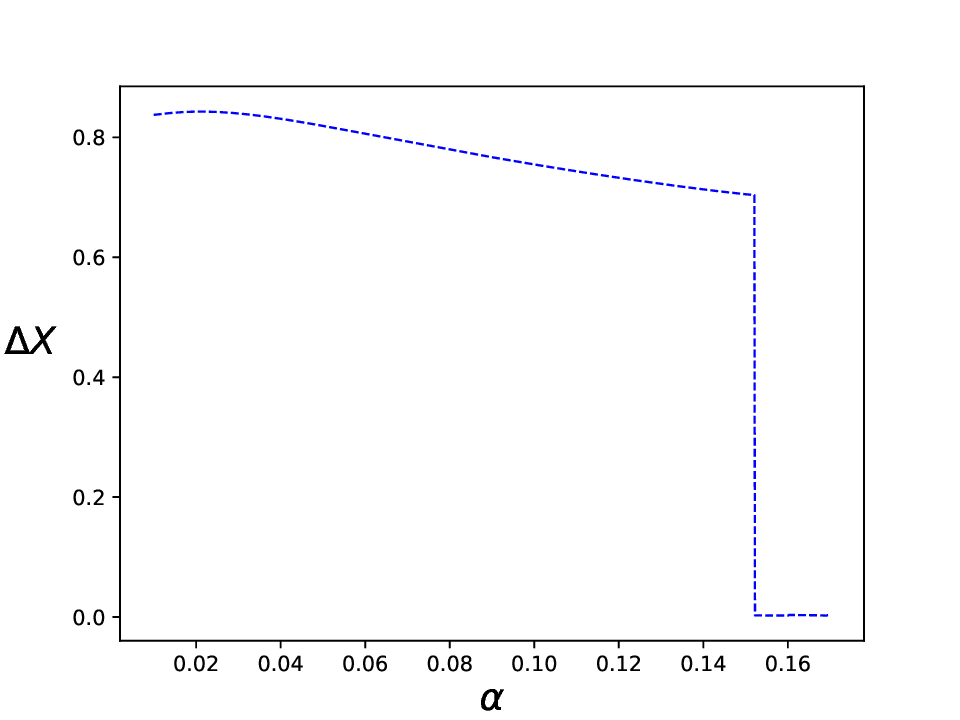}
        \caption{}
    \end{subfigure}
    \hfill
    \begin{subfigure}{0.49\textwidth}
        \centering
        \includegraphics[width=\linewidth]{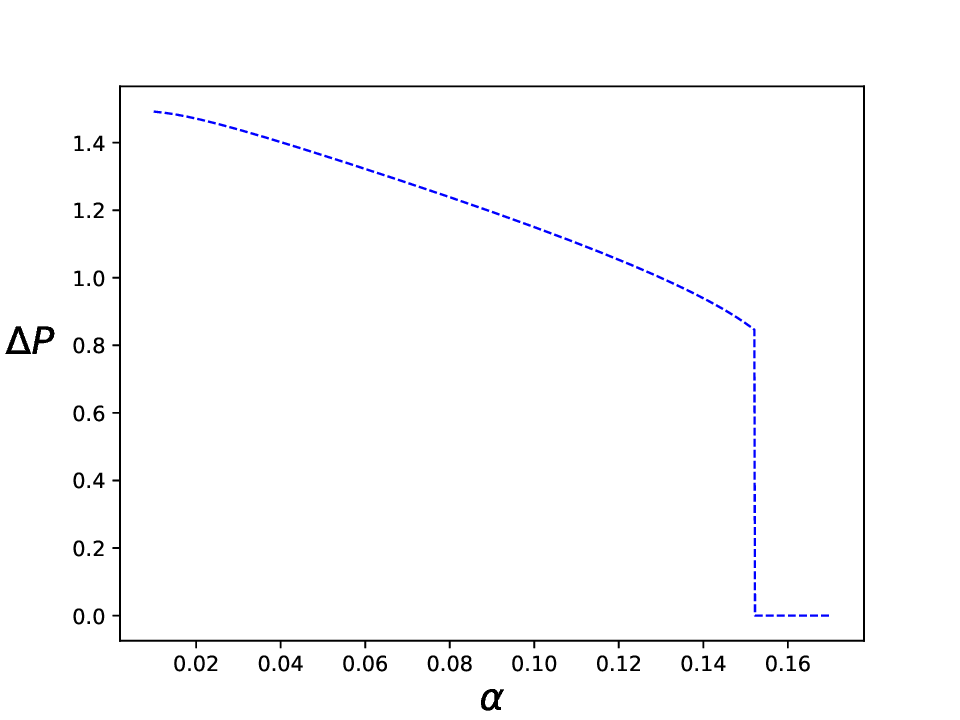}
        \caption{}
    \end{subfigure}

    \begin{subfigure}{0.5\textwidth}
        \centering
        \includegraphics[width=\linewidth]{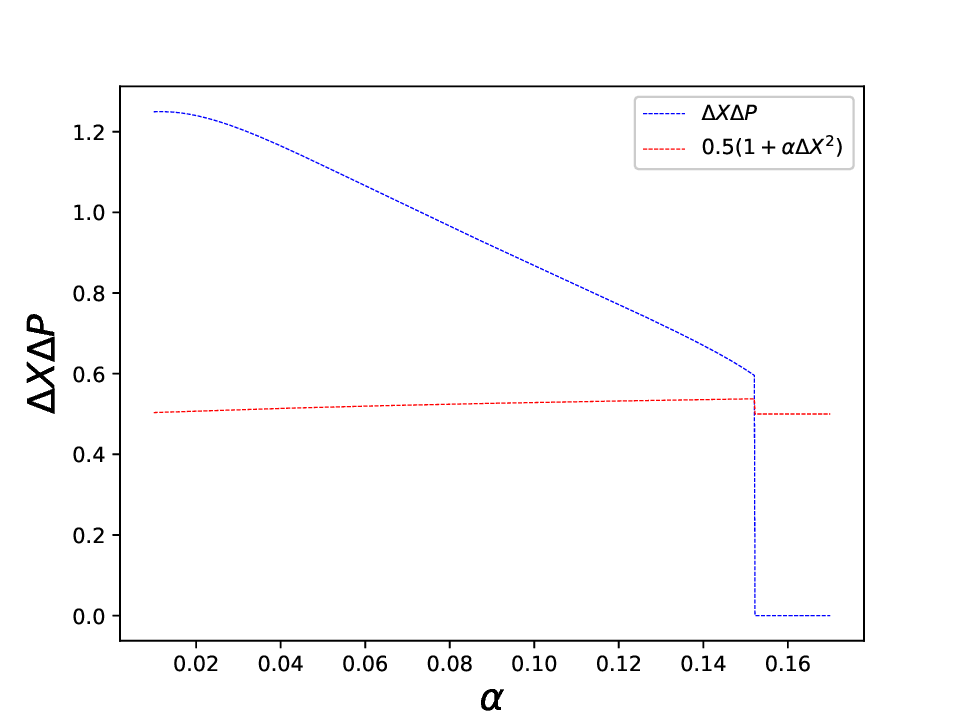}
        \caption{}

        \label{hoeup}
    \end{subfigure}

    \caption{ Variational solution of the GPE in the presence of HO potential: (a) $\Delta X$ vs $\alpha$ [Eq. (\ref{cc})]; (b) $\Delta P$ vs $\alpha$ [Eq. (\ref{momen})] \ ; (c) EUP validity: $\Delta X \Delta P$ against $0.5(1+\alpha \Delta X^2)$ [Inequality. (\ref{euppp})]}
    \label{harmeup}
\end{figure}

\begin{figure}[H]
\centering
  \begin{subfigure}{0.49\textwidth}
    \includegraphics[width=\linewidth]{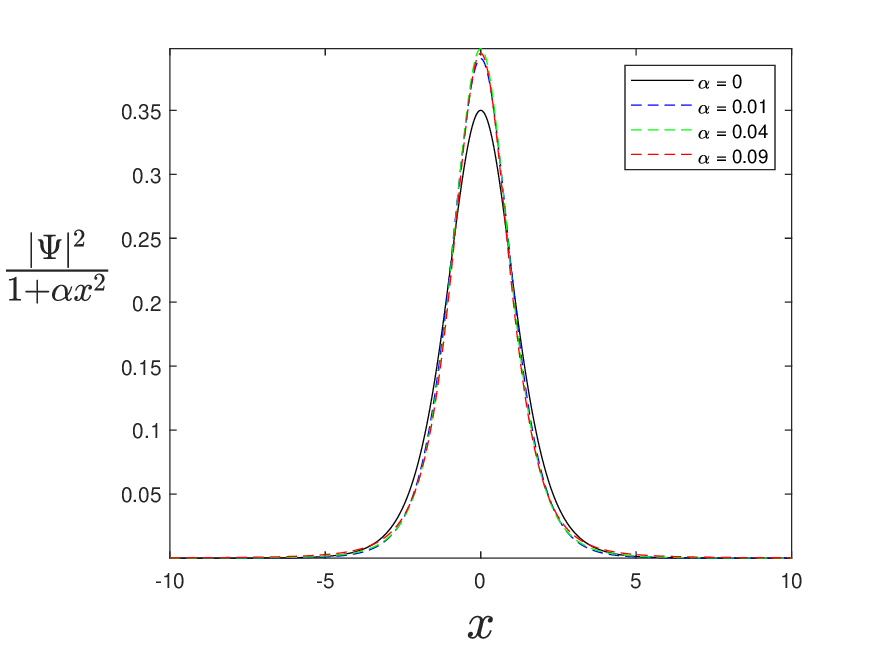}
    \caption{}
    \label{reva}
  \end{subfigure}
  \begin{subfigure}{0.49\textwidth}
    \includegraphics[width=\linewidth]{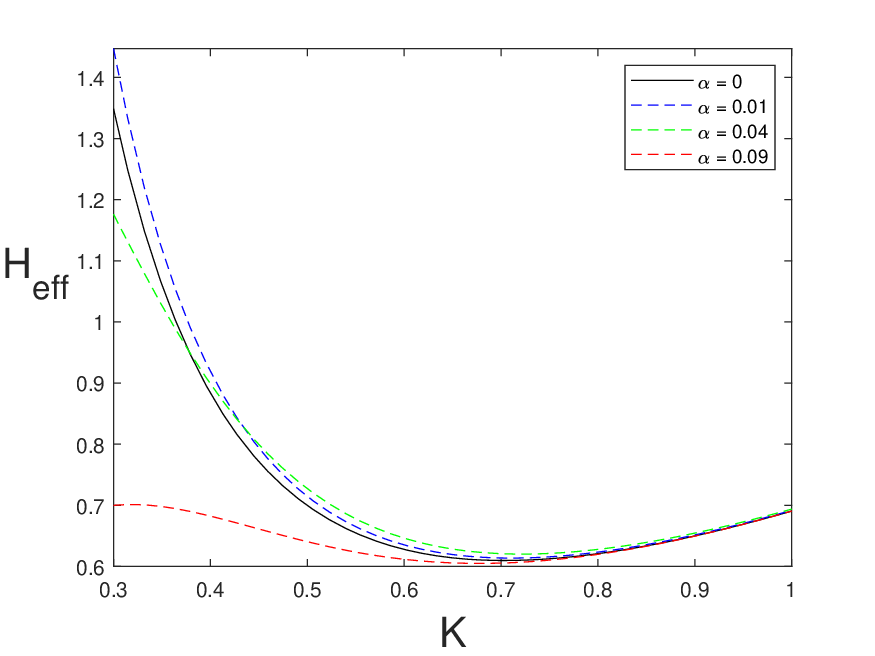}
    \caption{}
    
  \end{subfigure}
  \caption{(a) Probability density distribution corresponding to HO wave function described by [Eq. (\ref{ansatz})]. (b) Effective Hamiltonian as a function of $K$ with different values of $\alpha$ [Eq. (\ref{effectives})]}
   \label{ho}
\end{figure}

	 For the  HO potential $\displaystyle V(x)=\frac{x^{2}}{2}$ (where $\omega=1$), and utilizing the generalized momentum operator $\hat{P}=(1+\alpha x^{2})\hat{p}$, the deformed GPE takes the form
	
	\begin{gather}
		i\frac{\partial \tilde{\Psi}(\eta,t)}{\partial t}=-\frac{1}{2}\frac{\partial^{2} \tilde{\Psi}(\eta,t)}{\partial \eta^{2}}+g\left\arrowvert  \tilde{\Psi}(\eta,t) \right\arrowvert^{2} \tilde{\Psi}(\eta,t)+\frac{\tan^2(\eta\sqrt{\alpha})}{2\alpha}  \tilde{\Psi}(\eta,t). \label{harm}
	\end{gather}
Here, we used $x=\tan(\sqrt{\alpha}\eta)/\sqrt{\alpha}$. This transformation converts the HO potential to Pöschl–Teller (PT) potential which is considered a generalization of harmonic vibrators and gives phenomena such as anharmonicity \cite{Benkrane}. For the stationary solution, we use
\begin{gather}
 \tilde{\Psi}(\eta,t)=\phi(\eta) \exp(-i\mu t).
\end{gather}
Therefore, the effective classical field Hamiltonian for $g=-1$ is
\begin{equation}
		H_{eff}= \int \left[ \frac12\left\lvert \frac{\partial \phi(\eta)}{\partial \eta}\right\lvert^2 +\frac{  \tan^2(\sqrt{\alpha}\eta)}{2\alpha}  \lvert\phi(\eta)\lvert^2-\lvert\phi(\eta)\lvert^4\right] d\eta. 
		\label{effham}
	\end{equation} 
Now, we assume the following ansatz  
 \begin{gather}
 \phi(\eta)=\sqrt{\frac{K}{2}}\sech(K \eta), \label{ansatz}
 \end{gather}
$K$ being a variational parameter. Taking
into account the definitions  (\ref{effham}) and(\ref{ansatz}), the effective Hamiltonian can be expressed in terms of $K$ and $\alpha$ as 
\begin{gather}
H_{eff}= \frac18 K^3 J_1+\frac{1}{4\alpha} K J_2+ \frac14 K^2 J_3, \label{effectives}
\end{gather}
	Where
	
	\begin{equation}
		J_1=\int_{-\frac{\pi}{4\sqrt{\alpha}}}^{\frac{\pi}{4\sqrt{\alpha}}}\left(\sech^2(K\eta)-\sech^4(K\eta)\right)d\eta,
	\end{equation}
	
	\begin{equation}
		J_2=\int_{-\frac{\pi}{4\sqrt{\alpha}}}^{\frac{\pi}{4\sqrt{\alpha}}}\tan^2(\eta\sqrt{\alpha})\sech^4(K\eta)d\eta,
	\end{equation}
	and
	
	\begin{equation}
		J_3=\int_{-\frac{\pi}{4\sqrt{\alpha}}}^{\frac{\pi}{4\sqrt{\alpha}}}\sech^4(K\eta)d\eta.
	\end{equation}
	Similarly to the free case, we can compute $\Delta X$ and $\Delta P$ along with the validation of EUP as illustrated in Fig. \ref{harmeup}. Furthermore, by plotting the effective Hamiltonian as a function of $K$ for different values of $\alpha$ (see fig. \ref{ho}), we observe that Eq. (\ref{effham}) displays local minima at specific values of $K$. Inserting these values into the soliton ansatz yields profiles of the bright soliton depicted in Fig. \ref{reva}.

\subsubsection{Numerical analysis}

The numerical solutions of Eq. (\ref{harm}) for various values of $\alpha$ were obtained using SSFM \cite{Bao2003}. This method has proven to be a reliable approximation for the GPE \cite{Salman2014,Caliari2018,Wang2005}. The SSFM involves discretizing space and time and then alternating between Fourier transformation and applying evolution operators. The deformed GPE can be expressed as \cite{Bogomolov2006}

\begin{gather}
		i\frac{\partial \tilde{\Psi}(\eta,t)}{\partial t}= \left[\hat{\mathcal{L}}+\hat{\mathcal{N}}\right] \tilde{\Psi}(\eta,t), 
	\end{gather}
where the linear operator $\hat{\mathcal{L}}$ is defined as

\begin{equation}
    \hat{\mathcal{L}}=-\frac{1}{2}\frac{\partial^{2}}{\partial \eta^2}.
\end{equation}
The nonlinear operator $\hat{\mathcal{N}}$ is given by:

\begin{equation}
 \hat{\mathcal{N}} = -\left\arrowvert  \tilde{\Psi}(\eta,t) \right\arrowvert^{2} +\frac{\tan^2(\eta\sqrt{\alpha})}{2\alpha}.
\end{equation}
Generally, $\hat{\mathcal{L}}$ and $\hat{\mathcal{N}}$ operators do not commute. However, according to the Baker-Hausdorff formula, the error introduced by treating them as if they commute is of order $(\Delta t)^2$ \cite{Muslu2005,Sinkin2003,Musetti2018}. Taking small time steps $\Delta t$, we can approximate the solution by

\begin{equation}
    \Psi(\mathbf{\eta}, t + \Delta t) = e^{-i\hat{\mathcal{L}}\Delta t} e^{-i\hat{\mathcal{N}} \Delta t}  \Psi(\mathbf{\eta}, t).
\end{equation}

To apply the SSFM algorithm we first compute the solution considering the part that involves the nonlinear operator:

\begin{equation}
    \Psi(\mathbf{\eta}, t + \Delta t) = e^{-i \hat{\mathcal{N}}  \Delta t } \Psi(\mathbf{\eta}, t)
    \label{non}.
\end{equation}

\begin{figure}[htbp]
 \begin{center}

\begin{subfigure}[b]{0.49\textwidth}
\includegraphics[width=\textwidth]{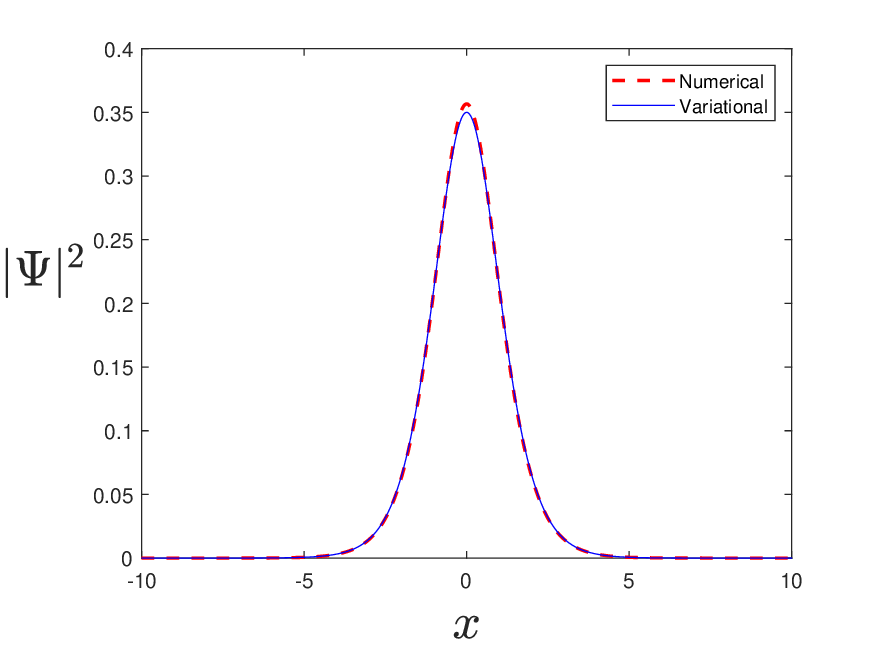}
\caption{}
\end{subfigure}%
  \hfill
  \begin{subfigure}[b]{0.49\textwidth}
    \includegraphics[width=\textwidth]{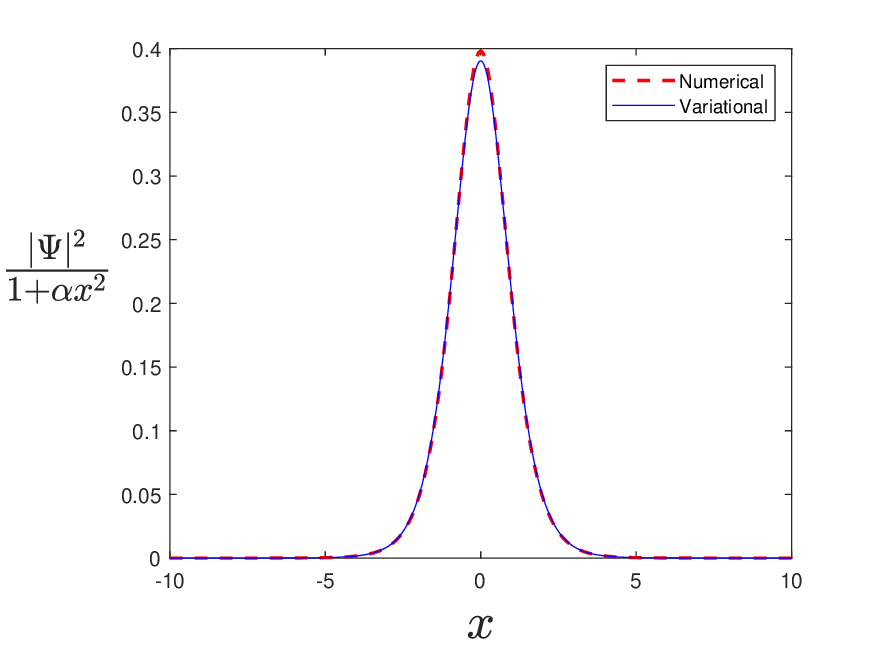}
    \caption{}
  \end{subfigure}%
  \hfill
  \begin{subfigure}[b]{0.49\textwidth}
    \includegraphics[width=\textwidth]{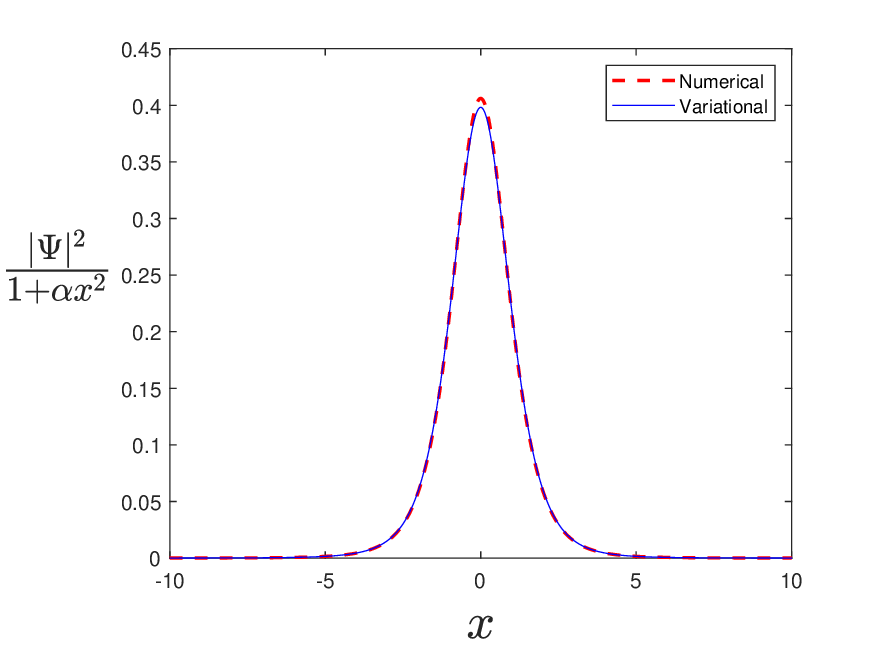}
    \caption{}
  \end{subfigure}%
  \hfill
  \begin{subfigure}[b]{0.49\textwidth}
    \includegraphics[width=\textwidth]{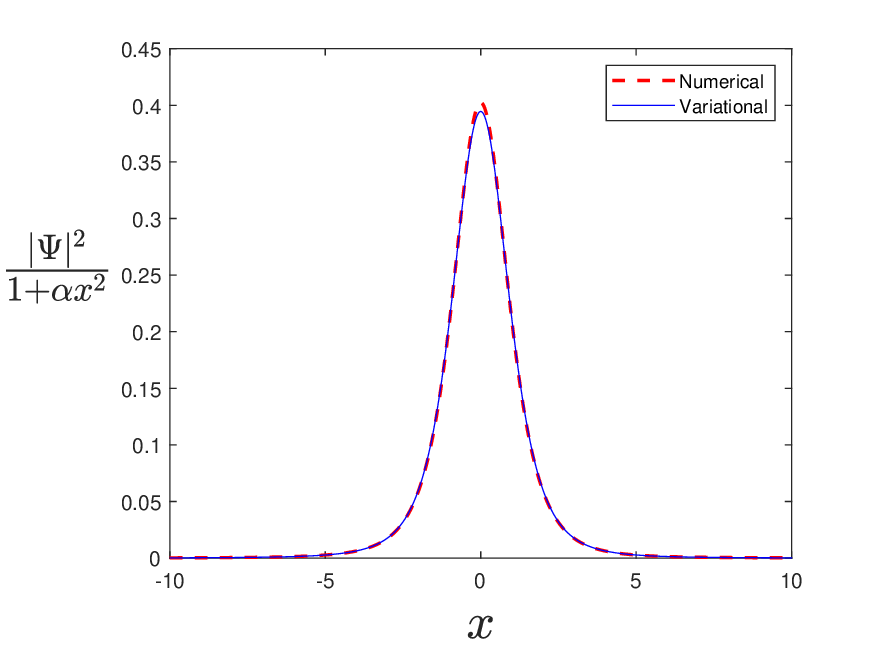}
    \caption{}
  \end{subfigure}

  \end{center}
  
  \caption{Numerical results compared to the variational solutions as function of $x$ for different values of $\alpha$: (a) $\alpha=0$, (b) $\alpha=0.01$, (c) $\alpha=0.04$, (d) $\alpha=0.09$}

  \label{numerical}
\end{figure}

In the next step, the solution (\ref{non}) undergoes a transformation into the frequency domain through the application of the Fourier transformation. Subsequently, it is advanced according to the linear operator, wherein the partial derivative is converted using $ik=i\frac{\partial}{\partial \eta}$ where $k$ is the wave number. Finally, the solution is subjected to an inverse Fourier transformation. This entire process is described by the expression 

\begin{equation}
    \Psi(\mathbf{\eta}, t + \Delta t) = \mathcal{F}^{-1} \left[ e^{-i \Delta t \left( -\frac12 \mathbf{k}^2\right)} \mathcal{F} \left[ e^{-i \hat{\mathcal{N}}  \Delta t } \Psi(\mathbf{\eta}, t) \right] \right],
\end{equation}
in which $\mathcal{F}$ and $\mathcal{F}^{-1}$ denote the forward and inverse Fourier transforms, respectively.
These steps are iteratively applied to solve Eq. (\ref{harm})  numerically. The code was written in MATLAB and we have taken 20000 Fourier modes with $\Delta t= 0.001$. The probability densities compared to the variational solutions are depicted in Fig. \ref{numerical}

\section{Discussions}
	
	We can observe the effects of  EUP on the BEC by plotting the probability densities, the position uncertainty  $\Delta X$ and the momentum uncertainty $\Delta P$ for different values of  $\alpha$. Based on the Figs. \ref{freebright}-\ref{ho} and \ref{numerical}, we give the following perspectives:
	\begin{enumerate}

       \item   In the examination of the free bright soliton solution, we have investigated  $\Delta X$ and $\Delta P$. The results, depicted in Fig. \ref{freebright} reveal that the EUP  does not apply for all values of the  $\alpha$. Consequently, it can be asserted that, within the framework of anti-de Sitter space, the bright soliton solution does not correspond to a physical state. 
		\item  The variations of $\Delta X$ and $\Delta P$ with EUP validation for a free dark soliton are shown in Fig. \ref{freedark}. It is evident that the curved space or the gravitational field enhances the precision of position measurements while concurrently diminishing the precision of momentum measurements. It is important to emphasize that these observations are right for the $\alpha$-values that make EUP valid and which can be extracted from Fig \ref{eupdark}. Specifically starting from $\alpha \simeq 0.3$, the inequality (\ref{euppp}) is fulfilled.
		\item  Fig. \ref{darksoliton} represents the probability density for a free dark soliton with position $x$ and for different values of $\alpha$. As the EUP impact increases, the amplitude grows and the width decreases and since the width physically is directly related to the $\Delta X$,  this confirms the decrease in the $\Delta X$ that we obtained in the Fig. \ref{darkxun}.
  \item  
  The variations of  $\Delta X$ and $\Delta P$ with EUP validation in terms of the $\alpha$ for the HO are illustrated in Fig. \ref{harmeup}. One can easily observe that as the $\alpha$ parameter increases, $\Delta X$ reaches a maximal value  after which the EUP causes a decrease in $\Delta X$. Meanwhile, the EUP directly reduces $\Delta P$  indicating a decrease in BEC depletion and an increase in BEC fraction with space-time curvature.
On the other hand, from $\alpha=0$ to $0.15<\alpha<0.155$, the EUP is valid  as depicted in Fig. \ref{hoeup}.

  \item   The probability density and effective Hamiltonian of the HO potential in the presence of EUP are presented in Fig. \ref{ho}. It is worth noting that the amplitude of probability increases with $\alpha$ and vice versa for the width. On the other hand, the effective Hamiltonian has a minimum value in terms of the variational parameter $K$ for a range of $\alpha \leq 0.15$ which reflects the stability of BEC. As $\alpha$ increases beyond this range, the effective Hamiltonian does not display local minima, indicating the instability of BEC.

  \item The variational solutions were validated using the SSFM and the two methods show remarkable agreement in Fig. \ref{numerical} with an error ranging between 0.074 and 0.077.

	\end{enumerate}

Furthermore, the quantum depletion in the context of BEC corresponds to the fraction of atoms that do not occupy the condensate at zero temperature due to correlation effects. However, the GPE mean-field approximation typically neglects the fluctuations exhibited by an interacting BEC \cite{Dalfovo1999,Fang2010}. Thus, to further investigate the effect of EUP on BEC depletion and fraction, it is more accurate to employ the Bogoliubov approximation, which incorporates quantum fluctuations into the description of the system \cite{Mller2012,Boudjema2022,Huang2006}.
	
	\section{Conclusion}	
	
	In this study, we explored the effects of the EUP on a BEC modeled by the deformed one-dimensional GPE. Specifically, we investigated free and harmonic trap scenarios. Our main findings suggest that the EUP significantly influences  $\Delta X$ and $\Delta P$ as well as the amplitude and the width of probability density. This effect varies according to $\alpha$. For instance, in the case of a dark soliton,  we found that the curved space increases the accuracy of measuring position and vice versa for the momentum of a free dark soliton. Additionally, we observed that the solution becomes normalizable under the effect of EUP in contrast to the ordinary case ($\alpha$=0), and the gravitational effect amplifies the amplitude of the probability density. 
 
 For the HO potential, we employed variational and numerical approaches and both approaches showed excellent agreement. The results indicate that the anti-de Sitter space leads to an increase in the amplitude of the probability density and a reduction in $\Delta X$ which contributes to a decrease in condensate depletion and thus an increase in the condensate fraction. Finally, it is important to note that these observations for null and HO potentials are valid for certain ranges of  $\alpha$, where the EUP inequality (\ref{euppp}) is fulfilled.

\section*{Data availability}

No new data were created or analysed in this study.

\section*{Acknowledgements}
We express our gratitude to two anonymous reviewers for their insightful remarks, which significantly enhanced the quality of the manuscript. Additionally, we would like to acknowledge the valuable exchange with Prof. Abdelaali Boudjemaa.
 
\bibliographystyle{unsrt} 
\bibliography{reference} 

\begin{thebibliography}{10}

\bibitem{31}
Weizhu Bao, Dieter Jaksch, and Peter~A Markowich.
\newblock Numerical solution of the gross--pitaevskii equation for bose--einstein condensation.
\newblock {\em Journal of Computational Physics}, 187(1):318--342, 2003.

\bibitem{32}
Franco Dalfovo, Stefano Giorgini, Lev~P Pitaevskii, and Sandro Stringari.
\newblock Theory of bose-einstein condensation in trapped gases.
\newblock {\em Reviews of modern physics}, 71(3):463, 1999.

\bibitem{GPE}
N~Bogoliubov.
\newblock On the theory of superfluidity.
\newblock {\em J. Phys}, 11(1):23, 1947.

\bibitem{GPE1}
Eugene~P Gross.
\newblock Hydrodynamics of a superfluid condensate.
\newblock {\em Journal of Mathematical Physics}, 4(2):195--207, 1963.

\bibitem{GPE2}
Lev~P Pitaevskii.
\newblock Vortex lines in an imperfect bose gas.
\newblock {\em Sov. Phys. JETP}, 13(2):451--454, 1961.

\bibitem{GPE4}
Aleksei Shabat and Vladimir Zakharov.
\newblock Exact theory of two-dimensional self-focusing and one-dimensional self-modulation of waves in nonlinear media.
\newblock {\em Sov. Phys. JETP}, 34(1):62, 1972.

\bibitem{GPE5}
A~Weller, JP~Ronzheimer, C~Gross, J~Esteve, MK~Oberthaler, DJ~Frantzeskakis, G~Theocharis, and PG~Kevrekidis.
\newblock Experimental observation of oscillating and interacting matter wave dark solitons.
\newblock {\em Physical review letters}, 101(13):130401, 2008.

\bibitem{GPE6}
Andrea Di~Carli, Craig~D Colquhoun, Grant Henderson, Stuart Flannigan, Gian-Luca Oppo, Andrew~J Daley, Stefan Kuhr, and Elmar Haller.
\newblock Excitation modes of bright matter-wave solitons.
\newblock {\em Physical Review Letters}, 123(12):123602, 2019.

\bibitem{GPE7}
Bettina Gertjerenken, Thomas~P Billam, Lev Khaykovich, and Christoph Weiss.
\newblock Scattering bright solitons: Quantum versus mean-field behavior.
\newblock {\em Physical Review A}, 86(3):033608, 2012.

\bibitem{GPE8}
L~Salasnich, A~Parola, and L~Reatto.
\newblock Modulational instability and complex dynamics of confined matter-wave solitons.
\newblock {\em Physical review letters}, 91(8):080405, 2003.

\bibitem{EPJ2020}
Mariusz~P Dabrowski and Fabian Wagner.
\newblock Asymptotic generalized extended uncertainty principle.
\newblock {\em The European Physical Journal C}, 80(7):676, 2020.

\bibitem{Physlett}
JR~Mureika.
\newblock Extended uncertainty principle black holes.
\newblock {\em Physics Letters B}, 789:88--92, 2019.

\bibitem{Merad2019}
A~Merad, M~Aouachria, M~Merad, and Tolga Birkandan.
\newblock Relativistic oscillators in new type of the extended uncertainty principle.
\newblock {\em International Journal of Modern Physics A}, 34(32):1950218, 2019.

\bibitem{Benkrane}
Abdelhakim Benkrane and Hadjira Benzair.
\newblock The thermal properties of a two-dimensional dirac oscillator under an extended uncertainty principle: path integral treatment.
\newblock {\em The European Physical Journal Plus}, 138(3):1--16, 2023.

\bibitem{Hamil2021}
B~Hamil, M~Merad, and Tolga Birkandan.
\newblock Pair creation in curved snyder space.
\newblock {\em International Journal of Modern Physics A}, 35(04):2050014, 2020.

\bibitem{Hamilscripta}
B~Hamil, M~Merad, and Tolga Birkandan.
\newblock The duffin-kemmer-petiau oscillator in the presence of minimal uncertainty in momentum.
\newblock {\em Physica Scripta}, 95(7):075309, 2020.

\bibitem{A.Merad2019}
A~Merad, M~Aouachria, and H~Benzair.
\newblock The eup dirac oscillator: A path integral approach.
\newblock {\em Few-Body Systems}, 61(4):36, 2020.

\bibitem{costafilho}
Raimundo~N Costa~Filho, Jo{\~a}o~PM Braga, Jorge~HS Lira, and Jos{\'e}~S Andrade~Jr.
\newblock Extended uncertainty from first principles.
\newblock {\em Physics Letters B}, 755:367--370, 2016.

\bibitem{gine}
Jaume Gin{\'e} and Giuseppe~Gaetano Luciano.
\newblock Modified inertia from extended uncertainty principle (s) and its relation to mond.
\newblock {\em The European Physical Journal C}, 80:1--8, 2020.

\bibitem{velocity}
E~Castellanos and JI~Rivas.
\newblock Planck-scale traces from the interference pattern of two bose-einstein condensates.
\newblock {\em Physical Review D}, 91(8):084019, 2015.

\bibitem{Qdeformed1}
Mahnaz Maleki, Hosein Mohammadzadeh, and Zahra Ebadi.
\newblock Nonextensive gross pitaevskii equation.
\newblock {\em International Journal of Geometric Methods in Modern Physics}, 20(12):2350216--330, 2023.

\bibitem{science1}
Frank Wilczek.
\newblock Anyons.
\newblock {\em Scientific American}, 264(5):58--65, 1991.

\bibitem{science2}
Frank Wilczek.
\newblock Disassembling anyons, 1992.

\bibitem{science3}
Behrouz Mirza and Hosein Mohammadzadeh.
\newblock Thermodynamic geometry of fractional statistics.
\newblock {\em Physical Review E}, 82(3):031137, 2010.

\bibitem{2013}
Manjun Ma and Zhe Huang.
\newblock Bright soliton solution of a gross--pitaevskii equation.
\newblock {\em Applied Mathematics Letters}, 26(7):718--724, 2013.

\bibitem{salazar}
Jesus Rogel-Salazar.
\newblock The gross--pitaevskii equation and bose--einstein condensates.
\newblock {\em European Journal of Physics}, 34(2):247, 2013.

\bibitem{Migenmi}
Salvatore Mignemi.
\newblock Extended uncertainty principle and the geometry of (anti)-de sitter space.
\newblock {\em Modern Physics Letters A}, 25(20):1697--1703, 2010.

\bibitem{Bolen}
Brett Bolen and Marco Cavaglia.
\newblock (anti-) de sitter black hole thermodynamics and the generalized uncertainty principle.
\newblock {\em General Relativity and Gravitation}, 37:1255--1262, 2005.

\bibitem{2016}
Manjun Ma, Chi Dang, and Zhe Huang.
\newblock Analytical expressions for dark soliton solution of a gross--pitaevskii equation.
\newblock {\em Applied Mathematics and Computation}, 273:383--389, 2016.

\bibitem{Lawson}
Lat{\'e}vi~M Lawson, Prince~K Osei, Komi Sodoga, and Fred Soglohu.
\newblock Path integral in position-deformed heisenberg algebra with maximal length uncertainty.
\newblock {\em Annals of Physics}, page 169389, 2023.

\bibitem{ReviewD}
Kourosh Nozari and Amir Etemadi.
\newblock Minimal length, maximal momentum, and hilbert space representation of quantum mechanics.
\newblock {\em Physical Review D}, 85(10):104029, 2012.

\bibitem{Bao2003}
Weizhu Bao, Dieter Jaksch, and Peter~A. Markowich.
\newblock Numerical solution of the gross–pitaevskii equation for bose–einstein condensation.
\newblock {\em Journal of Computational Physics}, 187(1):318–342, May 2003.

\bibitem{Salman2014}
Hayder Salman.
\newblock A time-splitting pseudospectral method for the solution of the gross–pitaevskii equations using spherical harmonics with generalised-laguerre basis functions.
\newblock {\em Journal of Computational Physics}, 258:185–207, February 2014.

\bibitem{Caliari2018}
M.~Caliari and S.~Zuccher.
\newblock Reliability of the time splitting fourier method for singular solutions in quantum fluids.
\newblock {\em Computer Physics Communications}, 222:46–58, January 2018.

\bibitem{Wang2005}
Hanquan Wang.
\newblock Numerical studies on the split-step finite difference method for nonlinear schr\"{o}dinger equations.
\newblock {\em Applied Mathematics and Computation}, 170(1):17–35, November 2005.

\bibitem{Bogomolov2006}
Ya.l. Bogomolov and A.D. Yunakovsky.
\newblock Split-step fourier method for nonlinear schrodinger equation.
\newblock In {\em DAYS on DIFFRACTION 2006}. IEEE, 2006.

\bibitem{Muslu2005}
G.M. Muslu and H.A. Erbay.
\newblock Higher-order split-step fourier schemes for the generalized nonlinear schr\"{o}dinger equation.
\newblock {\em Mathematics and Computers in Simulation}, 67(6):581–595, January 2005.

\bibitem{Sinkin2003}
O.V. Sinkin, R.~Holzlohner, J.~Zweck, and C.R. Menyuk.
\newblock Optimization of the split-step fourier method in modeling optical-fiber communications systems.
\newblock {\em Journal of Lightwave Technology}, 21(1):61–68, January 2003.

\bibitem{Musetti2018}
Simone Musetti, Paolo Serena, and Alberto Bononi.
\newblock On the accuracy of split-step fourier simulations for wideband nonlinear optical communications.
\newblock {\em Journal of Lightwave Technology}, 36(23):5669–5677, December 2018.

\bibitem{Dalfovo1999}
Franco Dalfovo, Stefano Giorgini, Lev~P. Pitaevskii, and Sandro Stringari.
\newblock Theory of bose-einstein condensation in trapped gases.
\newblock {\em Reviews of Modern Physics}, 71(3):463–512, April 1999.

\bibitem{Fang2010}
Shiang Fang, Ray-Kuang Lee, and Daw-Wei Wang.
\newblock Quantum fluctuations and condensate fraction during time-of-flight expansion.
\newblock {\em Physical Review A}, 82(3), September 2010.

\bibitem{Mller2012}
C~A M\"{u}ller and C~Gaul.
\newblock Condensate deformation and quantum depletion of bose–einstein condensates in external potentials.
\newblock {\em New Journal of Physics}, 14(7):075025, July 2012.

\bibitem{Boudjema2022}
Abdel\^aali Boudjem\^aa.
\newblock Weakly interacting bose gases with generalized uncertainty principle: Effects of quantum gravity.
\newblock {\em The European Physical Journal Plus}, 137(2), February 2022.

\bibitem{Huang2006}
Guoxiang Huang, L.~Deng, Jiaren Yan, and Bambi Hu.
\newblock Quantum depletion of a soliton condensate.
\newblock {\em Physics Letters A}, 357(2):150–153, September 2006.

\end{thebibliography}
\end{document}